# Time-Reversal Symmetry Breaking in the Fe-Chalcogenide Superconductors

N. Zaki,[1] G. Gu,[1] A. Tsvelik,[1] Congjun Wu[2]
and P.D. Johnson[1*]

1. CMPMSD, Brookhaven National Laboratory, Upton, NY 11973
2. Department of Physics, University of California, San Diego, California 92093, USA

**Topological superconductivity has been sought for in a variety of heterostructure systems, the interest being that a material displaying such a phenomenon could prove to be the ideal platform to support Majorana fermions, which in turn could be the basis for advanced qubit technologies. Recently the high Tc family of superconductors, $FeTe_{1-x}Se_x$, have been shown to exhibit the property of topological superconductivity and further, evidence has been found for the presence of Majorana fermions. We have studied the interplay of topology, magnetism and superconductivity in the $FeTe_{1-x}Se_x$ family using high-resolution laser-based photoemission. At the bulk superconducting transition, a gap opens at the chemical potential as expected. However, a second gap is observed to open at the Dirac point in the topological surface state. The associated mass acquisition in the topological state points to time-reversal symmetry breaking, probably associated with the formation of ferromagnetism in the surface layer. The presence of intrinsic ferromagnetism combined with strong spin-orbit coupling provides an ideal platform for a range of exotic topological phenomena.**

*Corresponding Author

Magnetism and superconductivity represent emergent ground states in condensed matter systems that often compete. In the high $T_c$ cuprates for example, the phase diagram is characterized by magnetism at low doping and superconductivity at higher doping levels.[1] There is some tendency for these two regions to overlap in the related Fe-based superconductors but note that the magnetic ground state is anti-ferromagnetic with neighboring spins anti-parallel as in the configuration associated with the Cooper pairs in spin-singlet superconductivity.[2] Ferromagnetism on the other hand with neighboring spins aligned parallel most definitely appears to counter the possibility of the normal spin alignment associated with superconductivity, at least for systems characterized by singlet pairing. However, the spin-orbit interaction associated with the formation of protected surface electron states on the surface of topological insulators can also play a role. A recent example being the demonstration that the high Tc superconductors, $FeTe_{1-x}Se_x$, supports topological surface states [3,4] reflecting the large spin-orbit interaction on the ligand Te atoms.[5] Indeed, in our own studies of this system we showed that spin-orbit effects [5] combined with the local moments [4] associated with the paramagnetic state resulted in an inverted gap at the zone boundary capable of supporting a topological state. In the present study, we examine the interaction between magnetism, superconductivity and topology in this fascinating and complex system. With the superconducting transition we observe a gap opening at the chemical potential, a characteristic of superconductivity reflecting the formation of Cooper pairs but with the same transition, we also find evidence of a second gap opening, now at the Dirac point associated with the topological surface state. The discovery of mass acquisition associated with the superconducting transition points to the breaking of a symmetry associated with the topological state. In fact with only a single cone at the center of the Brillouin zone the observation of mass acquisition leads to the conclusion that time reversal symmetry is broken, consistent with the development of some form of magnetic order. Ferromagnetism will break time reversal symmetry, anti-ferromagnetism will not. The observation of spontaneous time-reversal symmetry breaking may well bring helpful information to our understanding of the symmetry of the superconducting gap function in such systems.[6] The development of ferromagnetism is also highly suggestive that in the superconducting state this system could prove an ideal platform for the demonstration of a range of exotic topological phenomena. Indeed it has recently been suggested that in the presence of an external magnetic field topological superconductivity with a Dirac gap and the half quantum anomalous Hall state (hqah) can be competing ground states.[7] Further the same theory suggests that Majorana modes will exist on the boundary of such regions. It is possible that topological superconductors supporting such chiral edge modes could also exhibit a quantized thermal Hall conductance.

**Results:**

In figure 1(a) we show the photoemitted spectral intensity from $FeTe_{0.7}Se_{0.3}$ in the normal state. The intensity is obtained by summing the individual spectral intensities obtained with orthogonal light polarizations that effectively removes any matrix element effects. As in previous studies, the plot is characterized by two features, the bulk band dispersing downwards away from the $\Gamma-$ point and the topological state represented by the cone-like structure dispersing upwards from the Dirac point. Previous studies of the Dirac cone have demonstrated the helical spin structure associated with the state using either spin polarized photoemission with linearly polarized incident light[4] or photoemission with circularly polarized incident light.[5] In figure 1 we also show the photoemission spectra recorded from the material near the center of the zone as a function of the incident light polarization and temperature, corresponding in (b) and (c) to the normal state and in (d) and (e) to the superconducting state. In the latter spectra, the development of the peak associated with the superconducting transition is clearly visible for p-polarized light across the entire range of $k_{\parallel}$ measured. For s-polarized light, on the other hand, the onset of superconductivity is only visible in the vicinity of the topological state at the center of the zone. It is important to note that these studies benefit from the high spatial resolution, $< 20\mu m$, in that these samples display inhomogenities on length scales of $50\mu m$ or less.[8] In the appropriate geometry, matrix element effects in the photoemission process allow in principle, the identification of the orbital character of the initial state (SI Appendix, section 2). In the present case, the observation that the state associated with the superconducting transition is evident only with p-polarized light could suggest that the onset of superconductivity is orbital selective. Indeed in studies of a related system, $FeSe_{0.4}Te_{0.6}$, a Spectroscopic Imaging Scanning Tunneling Microscopy (SI-STM) study reported evidence for orbital ordering which the authors correlated with superconductivity.[9] Orbital ordering[10] or selectivity[11] has also been invoked in a separate SI-STM and associated theoretical studies of the related FeSe. Here we would comment that in our earlier study we proposed that spin-orbit coupling hybridizes the $d_{xz}$ and $d_{yz}$ orbitals in these materials. Thus the spin-orbit coupling results in two bands, $\alpha_1$ and $\alpha_2$, split by the energy of the coupling strength aligning orbital angular momentum and spin around the Γ point.[4] Thus the $\alpha_1$ band will have Kramers degenerate contributions given by $(d_{xz} + id_{yz})|\uparrow\rangle$ and $(d_{xz} - id_{yz})|\downarrow\rangle$ and the $\alpha_2$ band by $(d_{xz} + id_{yz})|\downarrow\rangle$ and $(d_{xz} - id_{yz})|\uparrow\rangle$.

We make one further observation relating to the peak associated with the superconducting state as observed in the spectrum taken with p-polarized light. As noted, it is observed across the entire zone center. This represents unusual behavior as pointed out in an earlier study of superconductivity in the material $Fe_{1+y}Se_xTe_{1-x}$, where the possibility of a Bose-Einstein Condensation (BEC) type transition as opposed to a BCS type transition was discussed.[12] The authors of that study noted that a BEC transition would result in a superfluid peak at $q = 0$ or the center of the zone as observed in the present study. A peak at the center of the zone contrasts with the BCS mechanism where we expect to see the Bogolyubov quasiparticles showing their peak intensity around $k_F$ away from the zone center. As noted by the authors of Ref. 10, the crossover from BCS to BEC behavior reflects the ratio of $\Delta/E_F$ where again $\Delta$ represents the superconducting gap and $E_F$ is the Fermi energy. The Bogolyubov dispersion associated with pairing in the topological state would be expected to disperse downwards at the Fermi wavevector crossing of ~0.03 Å$^{-1}$ and away from the zone center. The dispersion associated with pairing in the bulk $\alpha_2$ band could potentially initially disperse towards the zone center but the Fermi crossing for that band is much further out at approximately 0.15Å$^{-1}$.[5] We further note that evidence for BCS-BEC crossover behavior has also been reported in a recent study of the related FeSe system.[13] Here we simply note that the properties of the topological surface state at the center of the zone are highlighted by the use of incident s-polarized light and the properties of the bulk superconductivity away from the zone center by the use of incident p-polarized light We therefore chose to use s-polarized incident light to examine the properties of the topological state in detail.

Fig. 2 compares the temperature dependence of the photoemission spectra recorded along the surface normal, corresponding to $k_\parallel = 0$, using s-polarized light for temperatures from 20K down through the superconducting transition at $T_c$=14.5K to 6K, well into the superconducting state for (a) $FeTe_{0.55}Se_{0.45}$ and (b) $FeTe_{0.7}Se_{0.3}$. However there are notable differences between the two systems presented in Fig. 2. At 20K the increase in Te concentration on moving from $Te_{0.55}$ to $Te_{0.7}$ results in an increase in the intensity of the peak in the vicinity of the Dirac point, clearly showing the latter is related to the Te concentration. With the development of superconductivity, although more obvious in the 70% Te material, in both cases the two most prominent features in the spectra appear to be pushed apart as the temperature goes below the bulk $T_c$; the one at lower binding energy moving towards the chemical potential and the second at a binding energy beyond the Dirac point (located approximately 8.0 meV below the chemical potential as discussed in the SM) being pushed to higher binding energy. However, there are also differences between the peaks at low binding energy in the two systems. The peak in (a) is sharper with a width probably determined by the overall experimental resolution of 2.5 meV. The peak in (b) on the other hand is broader with a substructure suggesting the presence of more than one peak. Fig. 2(c) focuses specifically on the peak closer to $E_F$ in 2(b) over the same temperature range. The peak again continuously changes its structure, indicating different components, one appearing to have its development associated with the superconducting transition. This is particularly noticeable in the 10K spectrum where at least two peaks are clearly resolved.

The spectra in fig. 2 (a) could be interpreted simply as the development of a peak associated with the superconducting transition as is evidenced in many superconducting materials. Indeed in figure 3 we plot the temperature dependence of the peak observed at the Γ point for $FeTe_{0.55}Se_{0.45}$ as a function of temperature and compare that with the intensity of the superconducting peak at $k_F$. They are clearly both linked to the superconducting transition. However, the density of photoemission initial states at a given energy E in the vicinity of a gap has an energy dependence given by $E/(E^2-m^2)^{1/2}$ where 2m is the magnitude of the gap (SI Appendix, section 4). Thus the photoemission intensity peaks at the gap edge and decreases as one moves away from the gap. This is particularly evident in figure 1(e) where the TSS is most intense at the center of the zone and then falls off in intensity as the state disperses upwards towards the Fermi crossing at $k_F$. Figure 3 further confirms this picture. As we discuss below, these observations point to the formation of two gaps associated with the superconducting transition, one at the chemical potential, and a second gap at higher energies, namely at the Dirac point.

**Analysis and Conclusions**

In figure 4 we show a series of measured spectral intensities at the center of the zone, as a function of temperature above and below the superconducting transition. The spectral images clearly show the opening of a gap at the Dirac point at the transition temperature $T_c$. as is more evident in a video we present in the supplementary information (SI Appendix, Movie S1). In the latter showing the spectral response as a function of temperature from 20K down through $T_c$ to 5K, the opening of two gaps, one at the chemical potential and one at the Dirac point, at the superconducting transition is very obvious. However as we have discussed elsewhere, there is already a small gap at the Dirac point above $T_c$ in the normal state.[4] This gap may reflect superconducting fluctuations above $T_c$ [6] (SI Appendix, Section 6). Indeed NMR and transport measurements provide some evidence of a pseudogap associated with preformed pairs with an onset T* of 20K well above the Tc of 8K in the related material FeSe.[13] An alternative explanation would be the magnetic ordering of local moments in the surface layer as observed in a previous study of a doped topological insulator, also in the paramagnetic state in the bulk.[14]

As noted earlier, the development of a larger temperature dependent gap at the Dirac point and associated mass-acquisition, is an indicator of some form of time reversal symmetry breaking associated with the bulk superconducting transition. To investigate this further we start with the assumption that the observed electron spectrum at the surface has only one Fermi pocket. This is different from the bulk, where the Fermi surface has multiple sheets. Hence the surface quasiparticles can be described by $\psi_\sigma(\mathbf{k}), \psi_\sigma^+(\mathbf{k})$, where σ represents the index of a "Kramers doublet", or, an "effective spin" and **k,** the 2-dimensional momentum. Since photoemission shows well defined quasiparticle excitations, we treat them as surface modes not propagating into the bulk. Furthermore, the mass term for a single Dirac cone (Weyl fermion) breaks time reversal and therefore requires an internal magnetic field. Another important factor is the strong spin-orbit coupling, λ, caused presumably by the strong electric field on the surface (the Rashba effect). We therefore use a model that represents the continuum limit of the model adopted by Mascot *et al.* [15]

The standard approach to the description of superconductivity is to use the Nambu notation (SI Appendix, section 5). Thus we introduce the Nambu spinor $\Psi^T(k) = \left(c_\uparrow(k), c_\downarrow(k), c_\downarrow^+(-k), c_\uparrow^+(-k)\right)$, so that the Hamiltonian H is given by

$$H = \sum_k \widehat{\Psi}^+(k) \widehat{H}(k) \widehat{\Psi}(k) \qquad (1)$$

where the Hamiltonian can be written as

$$H(k) = \varepsilon(k)\tau^z \otimes I + \lambda\tau^z \otimes (k_y\sigma^x - k_x\sigma^y) + h\tau^z \otimes \sigma^z + (Re\Delta)\tau^x \otimes I + (Im\Delta)\tau^y \otimes I. \qquad (2)$$

Here the Pauli matrices $\tau^z$ act in the particle-hole space and the $\sigma^z$ matrices act in the spin space. $\varepsilon_0(k) = k^2/2m$, the bare dispersion, $\mu$ is the binding energy of the Dirac point and $h$ the Weiss field generated by the presumed ferromagnetic ordering in the surface region. Diagonalizing equation (2) results in the energy spectrum (SI Appendix, section 5):

$$E_\pm^2 = (\varepsilon_0(k) - \mu)^2 + \lambda^2 k^2 + h^2 + \Delta^2 k^2 \pm 2[(\varepsilon_0(k) - \mu)^2(\lambda^2 k^2 + h^2) + |\Delta|^2 \lambda^2 k^4]^{1/2} \quad (3)$$

In the vicinity of the Dirac point, well removed from the chemical potential equation (3) reduces to $E_\pm = (\varepsilon_0(k) - \mu) \pm \sqrt{\lambda^2 k^2 + h^2}$. Thus to get some idea of the magnitude of the magnetic field h we fit the measured dispersion in the vicinity of the Dirac point with the expression

$$E_\pm = \pm\sqrt{\lambda^2 k^2 + h^2}. \quad (4)$$

We could have simply made the assumption that the gap has the magnitude $2h$. However, we believe that the close proximity of the bulk bands to the bottom of the gap renders such an approach less accurate.

Fitting the measured spectral plots as shown in fig. 4(a) provides us with a measure of $h(T)$, the temperature dependence of the gap at the Dirac point. At the lowest temperatures the full opening of the gap is of the order of 8 meV. However, as noted earlier a gap of approximately 3.0 meV exists above Tc. We therefore associate the additional 5.0 meV with the development of superconductivity. By contrast, the full gap at the chemical potential directly associated with the cooper pairing in the superconducting state is of the order of 4.0 meV. This is to be compared with the full gap of 3.6 meV measured in the earlier ARPES study [3] and 3.8 meV measured in a recent STM study.[16] In Fig. 5 we compare the temperature dependence of a number of different gaps as a function of temperature. We also show the mean-field temperature dependence of the superconducting gap measured in the close lying bulk $\alpha_2$ hole band. The latter is out of range of the present laser study but $\Delta_0$ has been reported elsewhere. [17] In plotting $\Delta(T)$ for the bulk band we use the expression

$$\Delta(T) = \Delta_0 \tanh\left[\alpha\left(\frac{T_c}{T} - 1\right)^{\frac{1}{2}}\right] \quad (5)$$

where $\alpha$ is such that $\Delta_0 = \alpha k T_c$.[18] We are able to make a number of interesting observations from fig. 5. The temperature dependence of the gap associated with the Dirac point for both the $Te_{0.55}$ and the Te$_{0.7}$ samples follow the bulk superconductivity dependence. The superconducting gap in the topological state at the chemical potential for the Te$_{0.55}$ sample is identical in both the earlier study[3] and the present study and identical to the bulk superconductivity dependence.

Armed with the experimentally determined temperature dependence for the different gaps, we show in fig. 4(e-h) representative dispersions calculated from equation 3, to be compared with the experimentally observed behavior shown in fig. 4(a-d). We have not attempted to account for the 3 meV observed above Tc but rather show the opening of the gaps at the Dirac point and at the chemical potential with the onset of superconductivity. The agreement is again quite satisfying. We can make one further important observation regarding the plots shown in fig. 4(e-h). Elsewhere we made reference to the possibility that the gap observed above T$_c$ at the Dirac point could simply be due to sample misalignment.[4] However we note from fig. 4(e-h) that as we move to lower binding energy on the Dirac cone, away from the Dirac point, the influence of temperature change occurs at lower and lower temperatures. Further as we show (SI Appendix, section1) there is effectively zero movement of the sample as the temperature is lowered ruling out any temperature induced misalignment as the source of any change in gap magnitude.

Our study thus reveals a number of new and important observations that should stimulate further research and the push for even higher energy resolution. As noted earlier, the opening of a gap at the Dirac point, indicative of time-reversal symmetry breaking, points to the development of some form of ferromagnetic order in the surface region associated with the superconducting transition. Within the Ginsburg-Landau formulism TRSB requires mixing between two gap functions resulting in a complex order parameter (SI Appendix, section 6). Such a possibility has been discussed before although to date there has been no experimental verification. Through spin-orbit coupling the complex order parameter can induce spin magnetization as described in the SM and explained in detail elsewhere.[6] We note that alternative theories that may explain the appearance of the gap at the Dirac point have also been suggested.[19,20] The presence of ferromagnetism in the surface region associated with strong spin-orbit coupling would potentially make this system an ideal platform for supporting a range of unique topological phenomena.

**Materials and Methods:**

Single crystals of $FeTe_{1-x}Se_x$ were grown by a unidirectional solidification method. The nominal compositions had no excess Fe, and Te, measured by magnetic susceptibility, is 14.5 K in both samples. Single crystal samples were cleaved *in situ* at T≤ 5 K and base pressure of ≤$2x10^{-11}$ Torr. The photoemission studies were carried out using a 3 psec pulse width, 76 MHz rep rate Coherent Mira 900P Ti:Sapphire laser, the output of which was quadrupled to provide 6.0 eV incident light, focused into a spot on the sample ~ 20 μm in diameter. The polarization of the latter could be varied with the use of quarter and/or half wave plates to provide linear or circularly polarized light of arbitrary orientation on the Poincare sphere. Photoemission spectra were obtained using a Scienta SES 2002 electron spectrometer. The effective energy resolution is ~2.5meV (FWHM) as determined by the width of the sharpest features in the measured spectra. The angular resolution was ~ 0.002Å$^{-1}$ at the low photon energy used. The value of $E_F$ is determined by reference to a gold sample in contact with the FTS samples.

**Acknowledgements:** The authors acknowledge useful discussions with Igor Zalyznyak, John Tranquada, Dung-Hai Lee, Mike Norman, Lun-Hui Hu, Manfred Sigrist, Weiguo Yin, and Dirk Morr. Further the authors acknowledge excellent technical support from Fran Loeb. The work carried out at Brookhaven was supported in part by the U.S. DOE under Contract No. DE-AC02-98CH10886 and in part by the Center for Computational Design of Functional Strongly Correlated Materials and Theoretical Spectroscopy. C. W. at UCSD was supported by AFOSR FA9550-14-1-0168.

**Figure Captions:**

Fig. 1. **Photoemitted spectral intensities measured in the normal and superconducting states.** (a) Spectral intensity measured in the vicinity of the Γ-point ($k_\parallel = 0$) from FeTe$_{0.7}$Se$_{0.3}$. The total intensity corresponds to the sum of those intensities measured with incident p-polarized and s-polarized light. (b) Energy Distribution Curves measured in a region $\pm 0.08$Å around the Γ-point from FeTe$_{0.7}$Se$_{0.3}$ using p-polarized light and with the sample in the normal state at 20K. (c) The same as in (b) but now with s-polarized light. (d) and (e) are respectively the same as (b) and (c) but now with the sample held in the superconducting state at 6K.

Fig. 2. **Temperature dependence of the photoemission spectra along the surface normal.** (a) and (b) represent the temperature dependence from 20K to 6K of the measured spectra recorded along the surface normal from FeTe$_{0.55}$Se$_{0.45}$ and FeTe$_{0.7}$Se$_{0.3}$ respectively. (c) shows an expanded view of the temperature dependence of the peak immediately below the Fermi level for the spectra in (b).

Fig. 3. **Measured intensity in the Topological Surface State observed on FeTe$_{0.55}$Se$_{0.45}$ as a function of momentum.** The intensity of the TSS is compared at two points, Γ (maroon circles) and $k_F$ (blue triangles). In both cases the intensity is normalized to the intensity of the d-band at the center of the Brillouin zone. Note the difference in the intensity scales, $k_F$ on the left hand side and Γ on the right hand side.

Fig. 4. **Measured dispersions in the vicinity of the Γ-point as a function of temperature.** Spectral intensity measured in the vicinity of the Γ-point as a function of temperature. (a)-(d) correspond to 20K, 14K, 12K and 10K respectively. Superimposed over the measured intensities, the white dotted curves show the peak intensities, the red dashed curve shows the underlying Dirac cone and the solid red curves show the Dirac cone but now with a gap at the Dirac point determined by fitting with equation (4). (e)–(h) The calculated equivalent of (a)-(d) using the expression given in equation (3).

Fig. 5. **Temperature dependence of the different gap magnitudes.** Temperature dependence of the superconducting gap (red squares) from reference 3 and from the present study (purple diamonds) for the Te$_{0.55}$ system and the gap at the Dirac point from the present study for Te$_{0.55}$ (open circles) and Te$_{0.7}$ (blue circles) compared with the temperature dependence of the bulk superconductivity for the $\alpha_2$ band (dashed line) as determined in reference 17.

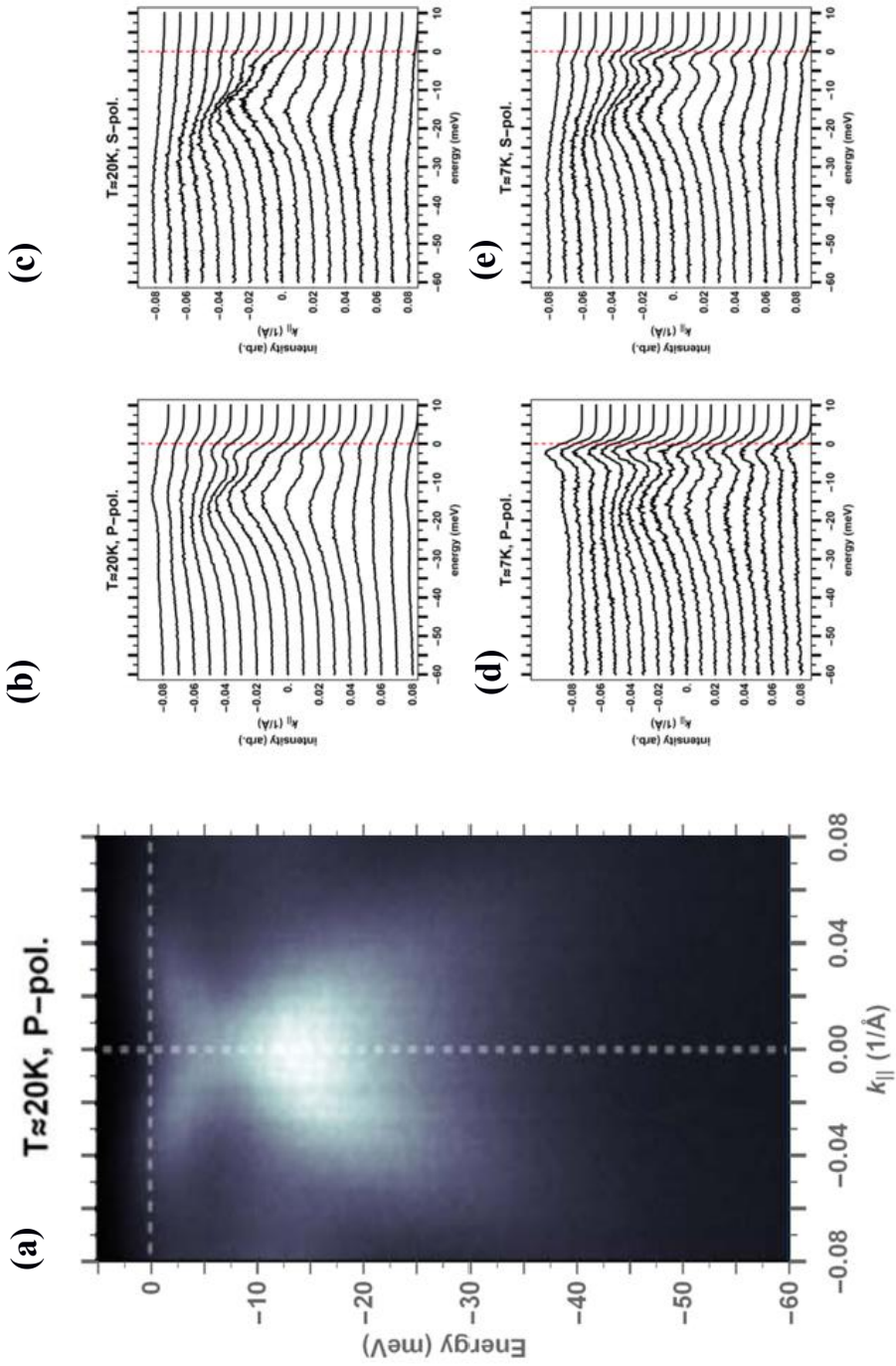

Fig. 1.

**Fig. 2.**

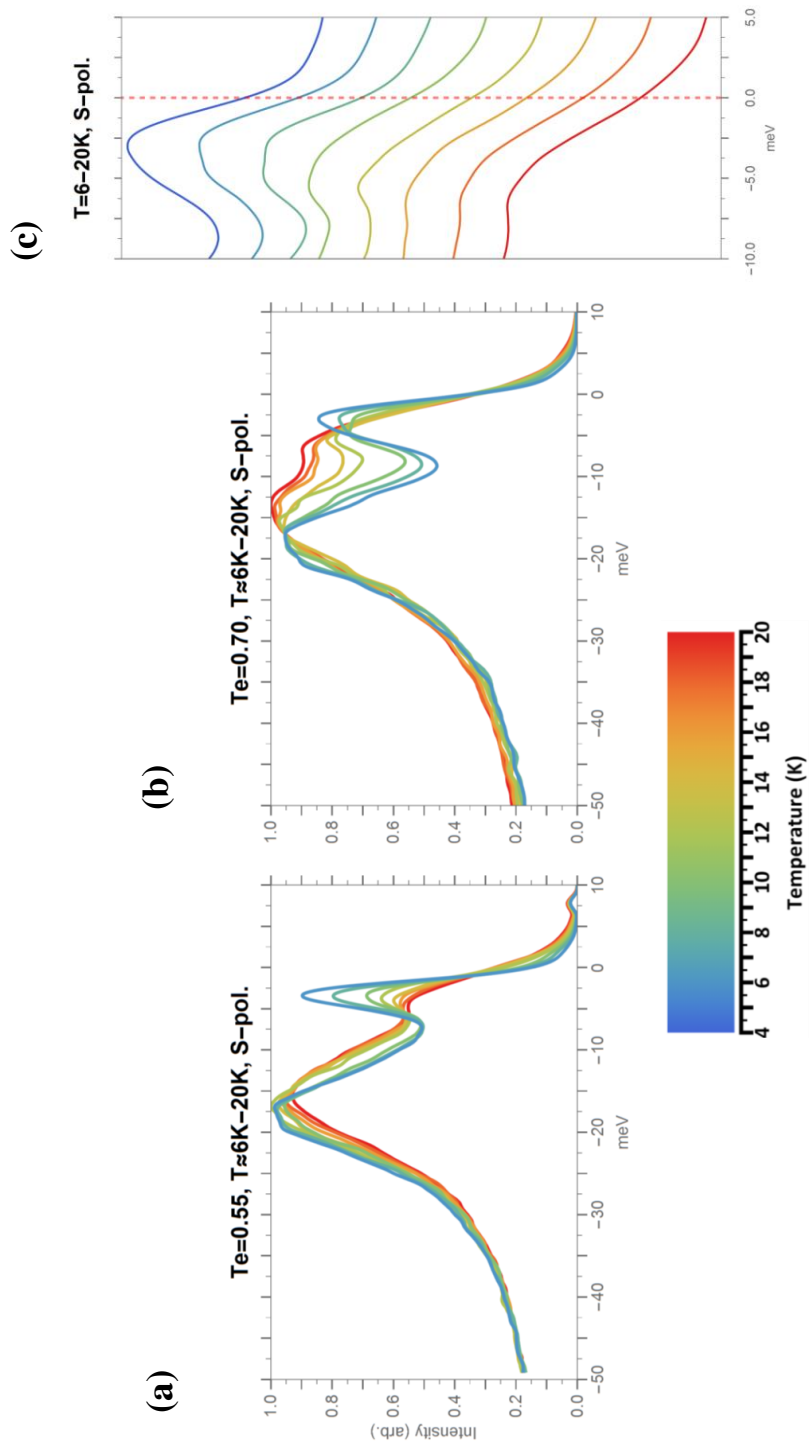

**Fig. 3**

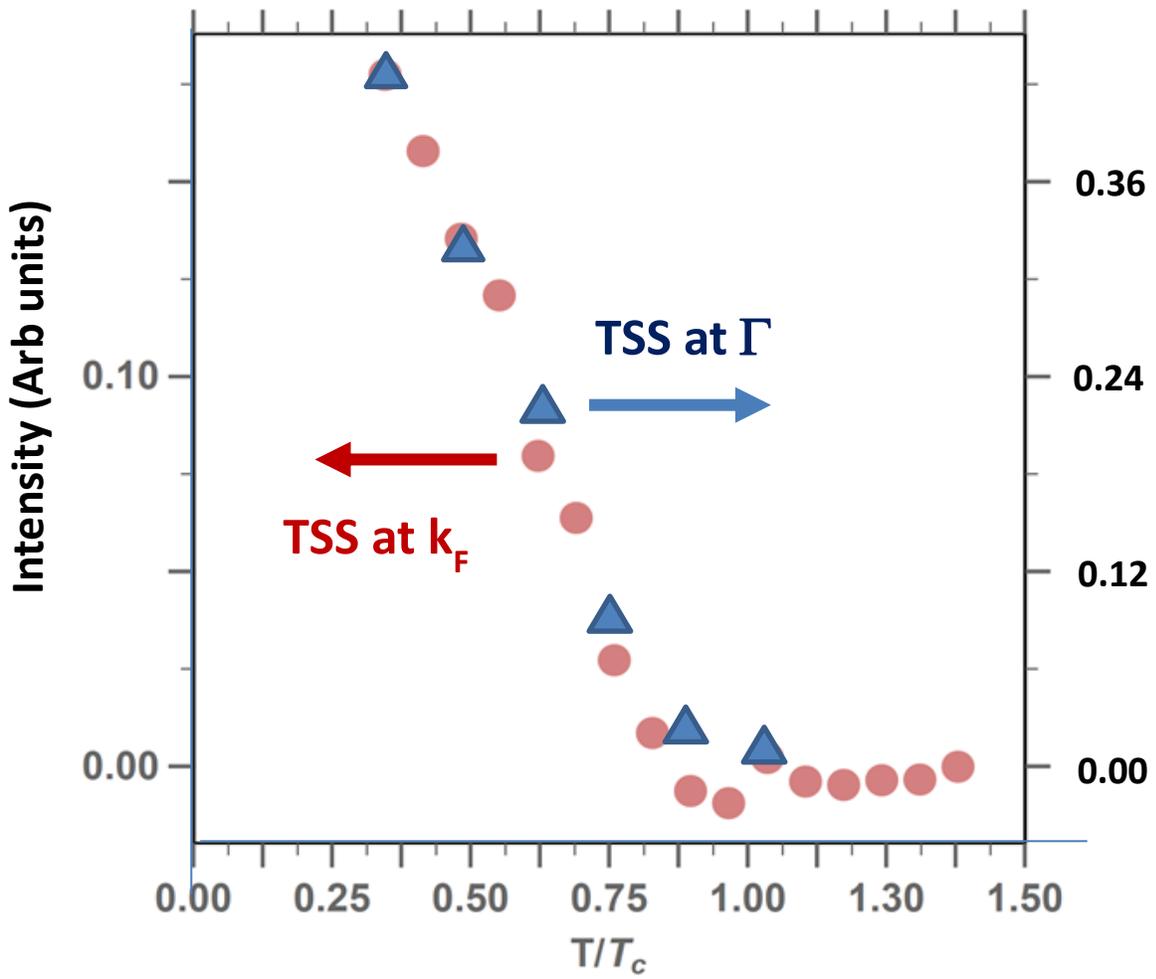

**Fig. 4.**

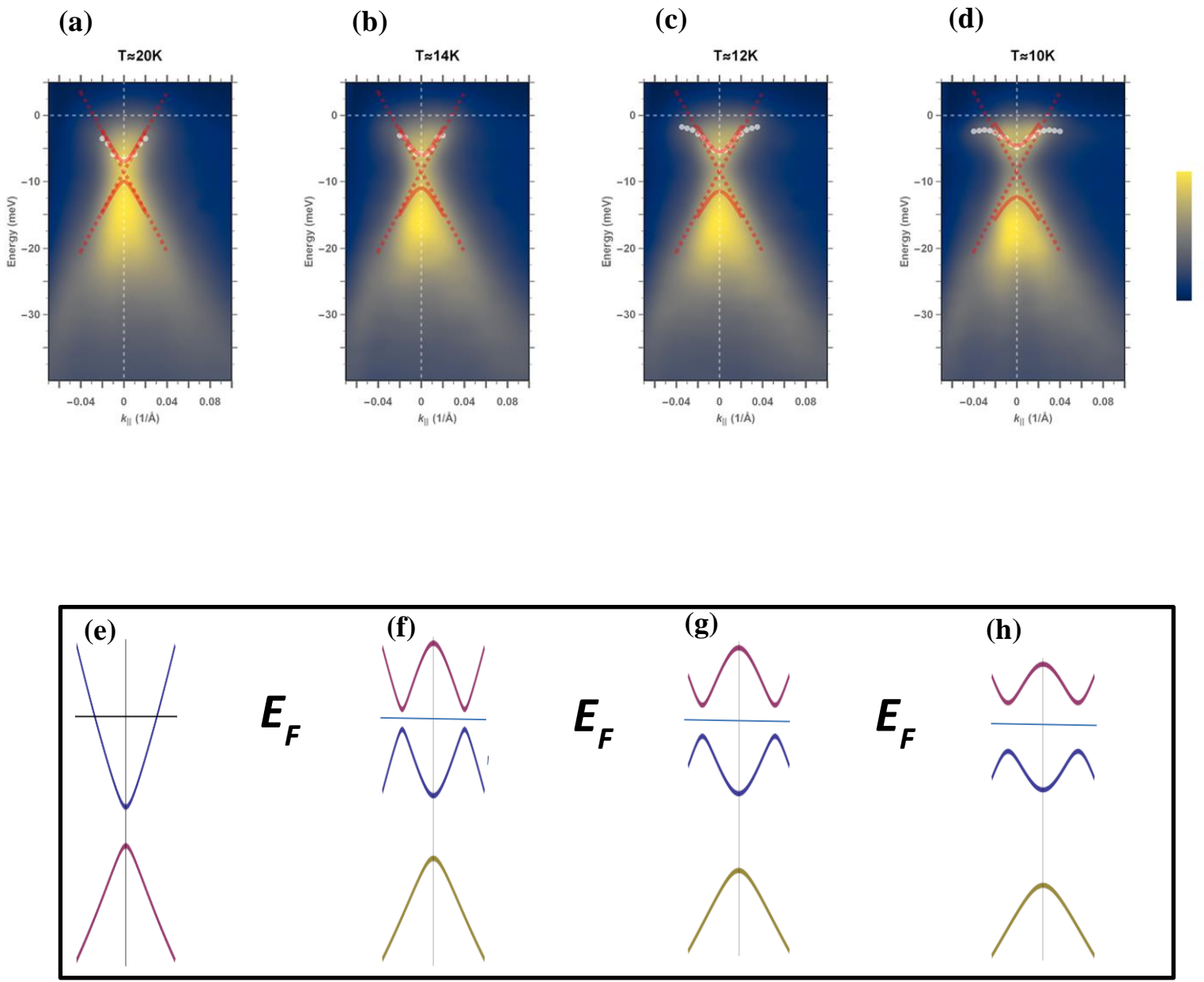

**Fig. 5.**

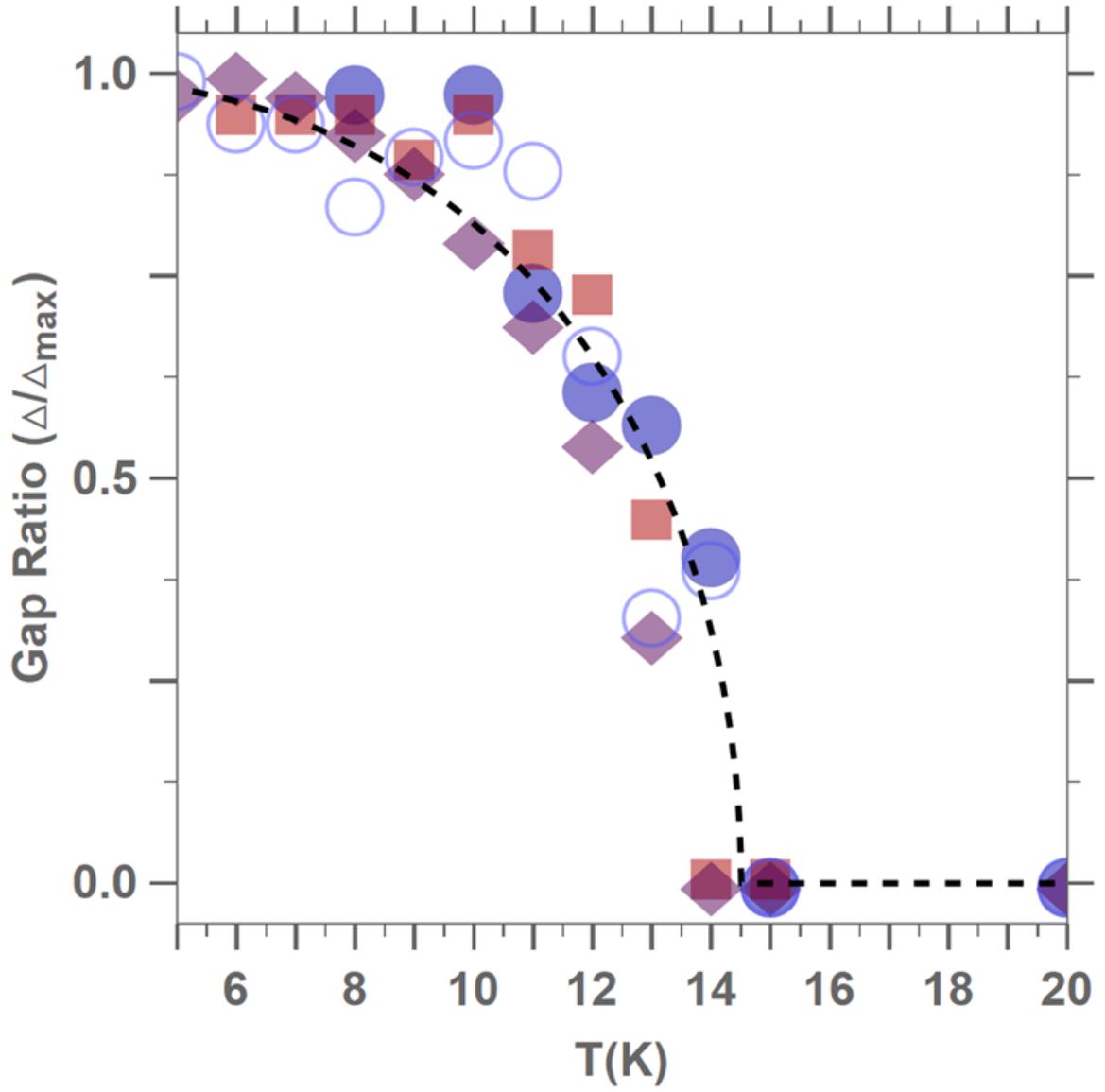